\documentclass[prb,twocolumn,showpacs,preprintnumbers,superscriptaddress,amsmath,amssymb]{revtex4}  % remove "%" sign for the PRB paper
\usepackage{graphicx}
\usepackage{dcolumn}
\usepackage{epstopdf} % am adaugat eu linia asta
\begin{document}

\title{First and second order magnetic and structural transitions in BaFe$_{2(1-x)}$Co$_{2x}$As$_{2}$}

\author{C. R. Rotundu}
\email[E-mail address: ]{CRRotundu@lbl.gov}
\affiliation{Materials Sciences Division, Lawrence Berkeley National Laboratory, Berkeley, CA 94720, USA}
%\author{N. E. Phillips}
%\affiliation{Materials Sciences Division, Lawrence Berkeley National
%Laboratory, Berkeley, CA 94720, USA}
%\affiliation{Department of Chemistry, University of California, Berkeley, California 94720, USA}
\author{R. J. Birgeneau}
\affiliation{Department of Physics, University of California, Berkeley, CA 94720, USA}
\affiliation{Department of Materials Science and Engineering, University of California, Berkeley, CA 94720, USA}
\date{\today}

\begin{abstract}

We present here high resolution magnetization measurements on  high-quality BaFe$_{2(1-x)}$Co$_{2x}$As$_{2}$, 0$\leq$x$\leq$0.046 as-grown single crystals. The results confirm the existence of a magnetic tricritical point in the ($x$,$T$) plane at x$^{m}_{tr}$$\approx$0.022 and reveal  the emergence of the heat capacity anomaly associated with the onset of the structural transition at x$^{s}$$\approx$0.0064. We show that the samples with doping near x$^{m}_{tr}$ do not show superconductivity, but rather superconductivity emerges at a slightly higher cobalt doping, x$\approx$0.0315.
\end{abstract}


\maketitle

The 122 series (AFe$_{2}$As$_{2}$, A = Ba, Sr, Ca, Eu) is one of the most studied among the newly discovered iron arsenide high temperature superconductors. One intriguing fact is that in this series superconductivity can be induced by doping in any of the three atomic sites; the cobalt doped system BaFe$_{2-2x}$Co$_{2x}$As$_{2}$ \cite{Sefat,Fisher} is, for instance, one of the most studied systems. The antiferromagnetic (spin-density wave) and structural (tetragonal to orthorhombic) transitions that are near-coincident in the parent compounds \cite{Rotter1} are concomitantly and gradually suppressed upon doping. A large number of papers have been written on the thermodynamic nature of the transitions in 122s arguing for either 1$^{st}$ order or 2$^{nd}$ order phase changes. If we solely consider the BaFe$_{2}$As$_{2}$ system, reports range from both 2$^{nd}$ order structural and magnetic phase transitions \cite{Rotter1}, to both transitions 2$^{nd}$ order but with the possibility of magnetic 1$^{st}$ order transition within 0.5 K of T$_{N}$ \cite{Wilson1}, to 1$^{st}$ of a order magnetic transition \cite{Kitagawa}, to indiscernibility between the two scenarios for the magnetic phase transition (did not see either the abrupt change at T$_{N}$ of the magnetic order parameter that is expected for a first-order transition, or the divergence of the correlation length at T$_{N}$ that would suggest a second-order transition) \cite{Matan}.

More recent combined high resolution X-ray diffraction and heat capacity measurements on exceptionally high quality BaFe$_{2}$As$_{2}$ crystals revealed a 1$^{st}$ order magnetic transition preceded by a structural transition that starts as a 2$^{nd}$ order transition at a slightly higher temperature but with a first order jump in the orthorhombic distortion coincident with the first order magnetic transition\cite{Rotundu1}.
Since data on some doped Ba122 samples show clear 2$^{nd}$ order magnetic and structural transitions \cite{Pratt,Nandi,Harriger}, it has been theoretically suggested that the magneto-structural transition in the parent is close to a tricritical point \cite{Giovannetti} that is tunable through doping. The only exception appears to be the case of the hole-doped Ba$_{1-x}$K$_{x}$Fe$_{2}$As$_{2}$ for which both the magnetic and structural transitions seem to be 1$^{st}$ order over the entire phase diagram \cite{Avci}. Further, ARPES measurements on cobalt doped Ba122 evidenced a Lifshitz transition at the onset of superconductivity \cite{Liu}, while high-resolution x-ray diffraction and x-ray resonant magnetic scattering (XRMS) pointed to a magnetic tricritical point at x$\approx$0.022, but with a structural transition whose onset is 2$^{nd}$ order across the whole cobalt doping range \cite{Kim}.

\begin{figure}[h]
%h=here, t=top, b=bottom, p=separate figure page
\begin{center}\leavevmode
\includegraphics[width=1.1\linewidth]{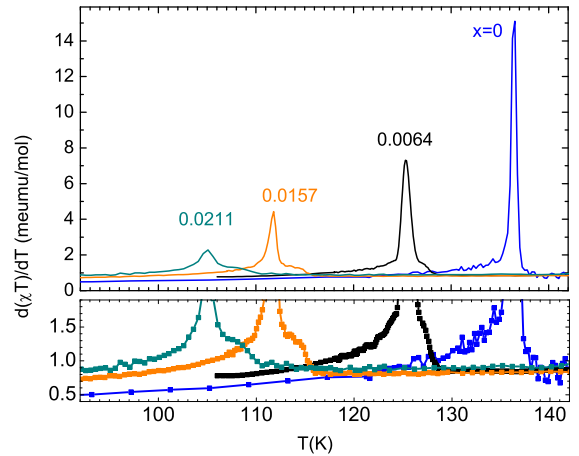}
\caption {d($\chi$T)/dT versus $T$ of the BaFe$_{2(1-x)}$Co$_{2x}$As$_{2}$, 0$\leq$x$\leq$0.0211 crystals. The higher temperature shoulder corresponds to the structural transition and the lower peak to the magnetic transition. The lower panel shows the same data near the ``base'' of the peaks on an expanded scale.}\label{fig2}\end{center}\end{figure}

\begin{figure}[h]
%h=here, t=top, b=bottom, p=separate figure page
\begin{center}\leavevmode
\includegraphics[width=1.1\linewidth]{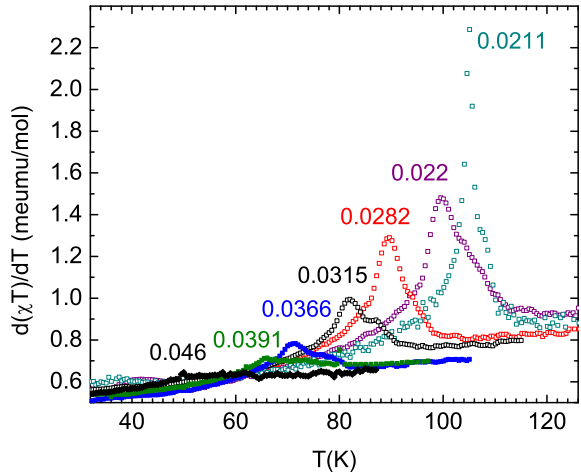}
\caption {d($\chi$T)/dT versus $T$ of the as-grown crystals BaFe$_{2(1-x)}$Co$_{2x}$As$_{2}$, 0.0211$\leq$x$\leq$0.046.}\label{fig3}\end{center}\end{figure}

\begin{figure}[h]
%h=here, t=top, b=bottom, p=separate figure page
\begin{center}\leavevmode
\includegraphics[width=1.1\linewidth]{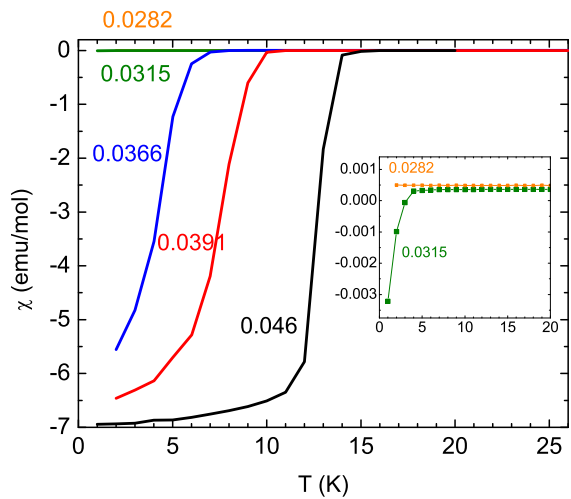}
\caption {Magnetic susceptibility $\chi$ versus $T$ of BaFe$_{2(1-x)}$Co$_{2x}$As$_{2}$ for cobalt doping x=0.0282, 0.0315, 0.0366, 0.091, and 0.046 measured in 20 Oe. The inset shows in magnified scale only the data of samples x=0.0282 and 0.0315.}\label{fig4}\end{center}\end{figure}

Magnetic and structural order parameters are normally obtained through scattering measurements. At the same time, measurements of thermodynamic quantities such as the heat capacity give direct information about the fluctuations associated with the order parameters and they can often be performed with quite high precision \cite{Dove}. For instance, fittings of high resolution heat capacity data in the critical region of a second order transition \cite{exponents} can give information about the values of the critical exponents and therefore on the dimensionality and symmetry of the systems under investigation. In the particular case of the 122s, because of the proximity of the magnetic and structural transitions, fits of the data near the transitions cannot be done reliably. Heat capacity measurements have been successfully used in the study of systems exhibiting doping driven $first$-$to$-$second$ order change of the transition \cite{Kozlowski1,Kozlowski2,Gallardo}. In these cases, high resolution heat capacity measurements, for instance, through a precise accounting of the latent heat \cite{latent}, can determine the  order of the transition extremely close to a tricritical point \cite{Cerro1,Romero}. However, these kinds of high resolution heat capacity measurements require a special setup and are very difficult to perform.

In this article we make use, rather, of the general theoretical argument of M. E. Fisher \cite{MEFisher} that shows that ``the variation of the magnetic specific heat of a $\emph{simple}$ antiferromagnet, in particular the singular behavior in the region of the transition, should be closely similar to the behavior of the function $\partial$($\chi$T)/$\partial$T, here $\chi$ is the zero-field susceptibility.'' The theoretical result had been successfully tested initially on MnO and MnF$_{2}$ \cite{MEFisher}. Although this theoretical result was initially meant for antiferromagnetic transitions, it seems that $\partial$($\chi$T)/$\partial$T mimics $C$ for both the magnetic and structural transitions of Co-doped Ba122 itself \cite{Fisher}, other 122s and 1111s as well. While establishing a precise equivalence between $C$(T) and $\partial$($\chi$T)/$\partial$T(T) is beyond the present study, we will exploit here the direct proportionality between the two physical measures at the transitions, i.e. \\$C$(T=T$_{t}$)$\propto$$\partial$($\chi$T)/$\partial$T(T=T$_{t}$), where T$_{t}$=T$_{N}$, T$_{s}$.

The measurements were made on as-grown single crystals of BaFe$_{2(1-x)}$Co$_{2x}$As$_{2}$ that were grown by the self-flux method \cite{Wang}. Inductively coupled plasma (ICP) and electron microprobe wavelength-dispersive X-ray spectroscopic (WDS) analysis were used to determine the actual stoichiometry of the samples, particular attention being given to the cobalt content. The samples (as determined by WDS) under study were: x=0, 0.0038$\pm$0.0006, 0.0064$\pm$0.0005, 0.0157$\pm$0.0007, 0.0211$\pm$0.0005, 0.0220$\pm$0.0005, 0.0282$\pm$0.0010, 0.0315$\pm$0.0011, 0.0366$\pm$0.0015, 0.0391$\pm$0.0011 and 0.046$\pm$0.0015. The $\pm$ represents the standard deviation from the average $x$ value of readings on ten randomly chosen points on each sample.
The magnetic susceptibility measurements on the samples were made using a Quantum Design Magnetic Property Measurement System (MPMS) in a magnetic field of 5 T parallel with the ($a$ $b$) crystallographic plane, unless otherwise noted. The machine was ``finely tuned'' \cite{QD} before the measurements to exploit the limits of its sensitivity, and also special care was taken to avoid oxygen contamination \cite{QD,Gregory}.

In brief, we describe high resolution magnetic susceptibility measurements on BaFe$_{2(1-x)}$Co$_{2x}$As$_{2}$, 0$\leq$x$\leq$0.04 that confirm the prediction of a magnetic tricitrical point at the doping x$\approx$0.022 and reveal in addition the emergence of the heat capacity anomaly associated with the onset of the second order structural transition at a lower doping, x$\approx$0.0064. Our data show further that the superconductivity emerges at a higher cobalt doping than the doping that corresponds to the magnetic tricritical point: at x$\approx$0.0315.

%Florin : Am mai pus eu niste $ $ care lipseau. Imi dadea eroare la compilare
Figures 1 and 2 show d($\chi$T)/dT versus $T$ of the as-grown crystals BaFe$_{2(1-x)}$Co$_{2x}$As$_{2}$ for 0$\leq$x$\leq$0.0211 and 0.0211$\leq$x$\leq$0.046, respectively. For the undoped sample (x=0) the corresponding d($\chi$T)/dT signature for both the antiferromagnetic (AFM) and structural transitions in this sample cannot be further apart than 0.25 K. This is consistent with previous results from our group.  Given the T$_{N}$ and T$_{s}$ dependence on the annealing (therefore synthesis) conditions \cite{Rotundu1}, it is not surprising that a (T$_{s}$-T$_{N}$) as large as 0.75 K has been reported \cite{Kim}.
Samples with x=0 and x=0.0038 (data not shown) show a single sharp peak which can be unambiguously identified with the combined magnetic and structural first order transition observed by Rotundu \emph{et al.} \cite{Rotundu1} and Kim \emph{et al.} \cite{Kim}. However, the anticipated heat capacity anomaly associated with the initial second order tetragonal-orthorhombic structural transition is not resolvable from the tail of the first order transition. However, as is evident in Fig. 1, for x=0.0064 cobalt doping, the heat capacity signature associated with the second order structural peak first emerges as a subtle but clear shoulder on the high temperature side of the first order jump (lower panel of Fig. 1).

Figure 3 shows magnetic susceptibility $\chi$ versus $T$ of BaFe$_{2(1-x)}$Co$_{2x}$As$_{2}$ for cobalt doping x=0.0282, 0.0315, 0.0366, and 0.0391 measured in 20 Oe. Samples x=0.022 and 0.0282 show a positive low temperature magnetization and therefore no signs of superconductivity. For sample x=0.0315, although $\chi$ shows a clear diamagnetic behavior, the superconducting volume fraction is lower than 1$\%$ (inset). The full superconducting volume fraction is reached with a $\approx$0.5$\%$ increase of $x$, i.e. at x$\approx$0.0366, attesting once more to the high quality of the crystals. The value of doping for which superconductivity is stabilized is in good agreement with values from the literature \cite{Ni}.

\begin{figure}[h]
%h=here, t=top, b=bottom, p=separate figure page
\begin{center}\leavevmode
\includegraphics[width=1.1\linewidth]{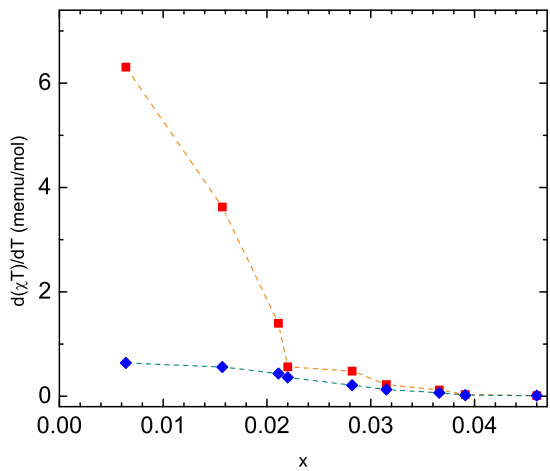}
\caption {The d($\chi$T)/dT magnitudes at T$_{N}$ ($\blacksquare$) and T$_{s}$ ($\blacklozenge$) of BaFe$_{2(1-x)}$Co$_{2x}$As$_{2}$ versus $\emph{x}$, for 0.0064$\leq$x$\leq$0.046.}\label{fig4}\end{center}\end{figure}

\begin{figure}[h]
%h=here, t=top, b=bottom, p=separate figure page
\begin{center}\leavevmode
\includegraphics[width=1.1\linewidth]{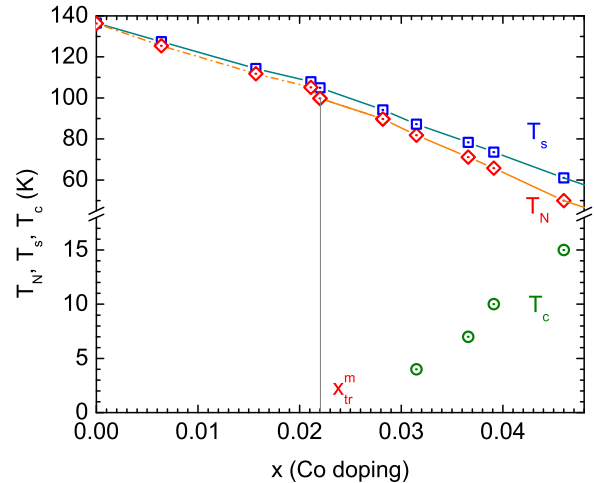}
\caption {The phase diagram for the doping range near the tricritical points. For T$_{s}$ and T$_{N}$ the discontinuous line indicates first order transition and the continuous line second order transition.}\label{fig4}\end{center}\end{figure}

%Florin : Am mai pus eu niste $ $ care lipseau. Imi dadea eroare la compilare
Figure 4 shows the magnitudes of d($\chi$T)/dT at T=T$_{N}$ ($\blacksquare$) and at T=T$_{s}$ ($\blacklozenge$) of BaFe$_{2(1-x)}$Co$_{2x}$As$_{2}$ versus $\emph{x}$ for 0.0064$\leq$x$\leq$0.0391. The x=0 point is not included in the graph since the peak associated with the onset second order structural transition could not be separated from the large first order jump. For d($\chi$T)/dT(T=T$_{N}$) the crossover points x$_{tr}$ marks the separation between a globally abrupt and non-monotonic doping evolution (characteristic of 1$^{st}$ order transitions region) and a monotonic variation (region characteristic of 2$^{nd}$ order transitions). The magnetic crossover at x$^{m}_{tr}$$\approx$0.022, appears to be a tricritical point; this confirms, albeit with much more precision, the suggestion of Kim \emph{et al.} \cite{Kim}. Any possible structural crossover is more difficult to discuss given that at x=0 the transition is characterized by a second order onset followed by a large first order jump presumably driven by the first order magnetic transition. At x$^{s}$$\approx$0.0064 the heat capacity of the second order structural transition becomes clearly visible.
Figure 5 summarize the above results in a phase diagram ($T$ $x$). For T$_{s}$ and T$_{N}$ the discontinuous line indicates first order transition and the continuous line second order transition. The phase diagram here is similar to the one for the 122s predicted by Cano $\emph{et al.}$ \cite{Cano}. Their Ginzburg-Landau model contains a magnetoelastic coupling term as a key ingredient. Using a slight variation of the afore mentioned model, Kim et al. \cite{Kim} predicted that the magnetic tricritical point is x$\approx$0.022. It should be noted that Wilson \emph{et al.} \cite{Wilson1} have shown that in BaFe$_{2}$As$_{2}$ below the first order transition the magnetic and structural order parameters exhibit identical temperature dependencies. This requires that there is a dominant biquadratic coupling between the magnetic and structural order parameters. The biquadratic term is not included in the above-mentioned theories.

In summary we have systematically studied the magnetic susceptibility of high quality BaFe$_{2(1-x)}$Co$_{2x}$As$_{2}$, 0$\leq$x$\leq$0.046 as grown single crystals. Our measurements confirm the existence of a magnetic tricritical point at the cobalt doping x$\approx$0.022. They also demonstrate that the anomaly associated with the putative second order structural transition emerges clearly separated from the first order combined magnetic and structural transition at a doping of about 0.0064. We show further that the superconductivity emerges at higher cobalt doping than that at the magnetic tricritical point: namely x$\approx$0.0315.

We thank Dung-Hai Lee, Alan I. Goldman, and J\"{o}rg Schmalian for valuable communications about their results and to Edith Bourret and John Heron for assistance. This work was supported by the Director, Office of Science, Office of Basic Energy Sciences, U.S. Department of Energy, under Contract No. DE-AC02-05CH11231 and Office of Basic Energy Sciences US DOE DE-AC03-76SF008.


\begin{thebibliography}{34}

\bibitem{Sefat}
A. S. Sefat, R. Jin, M. A. McGuire, B. C. Sales, D. J. Singh, and D. Mandrus, Phys. Rev. Lett. {\bf 101}, 117004 (2008).

\bibitem{Fisher}
J.-H. Chu, J. G. Analytis, C. Kucharczyk, and I. R. Fisher, Phys. Rev. B {\bf 79}, 014506 (2009).

\bibitem{Rotter1}
M. Rotter, M. Tegel, D. Johrendt, I. Schellenberg, W. Hermes, and R. P\"{o}ttgen, Phys. Rev. B {\bf 78}, 020503(R) (2008).

\bibitem{Wilson1}
S. D. Wilson, Z. Yamani, C. R. Rotundu, B. Freelon, E. Courchesne, and R. J. Birgeneau, Phys. Rev. B {\bf 79}, 184519 (2009).

\bibitem{Kitagawa}
K. Kitagawa, N. Katayama, K. Ohgushi, M. Yoshida, and M. Takigawa, J. Phys. Soc. Jpn. {\bf 77}, 114709 (2008).

\bibitem{Matan}
K. Matan, R. Morinaga, K. Iida, and T. J. Sato, Phys. Rev. B {\bf 79}, 054526 (2009).

\bibitem{Rotundu1}
C. R. Rotundu, B. Freelon, T. R. Forrest, S. D. Wilson, P. N. Valdivia, G. Pinuellas, A. Kim, Z. Islam, J. -W. Kim, E. Courchesne, N. E. Phillips, and R. J. Birgeneau, Phys. Rev. B {\bf 82}, 144525 (2010).

\bibitem{Pratt}
D. K. Pratt, W. Tian, A. Kreyssig, J. L. Zarestky, S. Nandi, N. Ni, S. L. Bud'ko, P. C. Canfield, A. I. Goldman, and R. J. McQueeney, Phys. Rev. Lett. {\bf 103}, 087001 (2009).

\bibitem{Nandi}
S. Nandi, M. G. Kim, A. Kreyssig, R. M. Fernandes, D. K. Pratt, A. Thaler, N. Ni, S. L. Bud'ko, P. C. Canfield, J. Schmalian, R. J. McQueeney, and A. I. Goldman, Phys. Rev. Lett. {\bf 104}, 057006 (2010).

\bibitem{Harriger}
L. W. Harriger, A. Schneidewind, S. Li, J. Zhao, Z. Li, W. Lu, X. Dong, F. Zhou, Z. Zhao, J. Hu, and P. Dai, Phys. Rev. Lett. {\bf 103}, 087005 (2009).

\bibitem{Giovannetti}
G. Giovannetti, C. Ortix, M. Marsman, M. Capone, J. van den Brink, and J. Lorenzana, arXiv:1009.0009v1 (unpublished).

\bibitem{Avci}
S. Avci, O. Chmaissem, E. A. Goremychkin, S. Rosenkranz, J.-P. Castellan, D. Y. Chung, I. S. Todorov, J. A. Schlueter, H. Claus, M. G. Kanatzidis, A. Daoud-Aladine, D. Khalyavin, and R. Osborn, arXiv:1102.1933 (unpublished).

\bibitem{Liu}
C. Liu, T. Kondo, R. M. Fernandes, A. D. Palczewski, E. D. Mun, N. Ni, A. N. Thaler, A. Bostwick, E. Rotenberg, J. Schmalian, S. L. Bud'ko, P. C. Canfield and A. Kaminski, Nature Physics {\bf 6}, 419 (2010).

\bibitem{Kim}
M. G. Kim, R. M. Fernandes, A. Kreyssig, J. W. Kim, A. Thaler, S. L. Bud'ko, P. C. Canfield, R. J. McQueeney, J. Schmalian, and A. I. Goldman, Phys. Rev. B {\bf 83}, 134522 (2011).

\bibitem{Dove}
M. T. Dove, Structure and Dynemics -- An atomic view of materials, Oxford University Press, Cambridge, UK (2010).

\bibitem{exponents}
Critical exponents are not defined at a first-order transition.

\bibitem{Kozlowski1}
A. Koz{\l}owski, Z. K\c{a}kol, R. Zalecki, K.S. Knight, and J.M. Honig, J. Phys. IV France {\bf 7}, 591 (1997).

\bibitem{Kozlowski2}
A. Koz{\l}owski, Z. K\c{a}kol, R. Zalecki, K. S. Knight, J. Sabol and J. M. Honig, J. Phys.: Condens. Matter {\bf 11}, 2749 (1999).

\bibitem{Gallardo}
M. C. Gallardo, F. J. Romeo, S. A. Hayward, E. K. Salje, and J. del Cerro, Mineralogical Magazine {\bf 64}, 971 (2000).

\bibitem{latent}
The latent heat is the thermodynamic measure that determines the first-order character of a phase transition.

\bibitem{Cerro1}
J. del Cerro, F.J. Romero, M.C. Gallardo, S.A. Hayward, J. Jim\'{e}nez, Thermochim. Acta {\bf 343}, 89 (2000).

\bibitem{Romero}
F. J. Romero, M. C. Gallardo, J. Jim\'{e}nez, J. del Cerro, Thermochim. Acta {\bf 372}, 25 (2001).

\bibitem{MEFisher}
M. E. Fisher, Philosophical Magazine {\bf 7}, 1731 (1962).

\bibitem{Wang}
X. F. Wang, T. Wu, G. Wu, H. Chen, Y. L. Xie, J. J. Ying, Y. J. Yan, R. H. Liu, and X. H. Chen, Phys. Rev. Lett. {\bf 102}, 117005 (2009).

\bibitem{QD}
MPMS Application Note 1014-210B, 03-04-1997.

\bibitem{Gregory}
S. Gregory, Phys. Rev. Lett. {\bf 40}, 723 (1978).

\bibitem{Ni}
N. Ni, M. E. Tillman, J.-Q. Yan, A. Kracher, S. T. Hannahs, S. L. Bud'ko, and P. C. Canfield, Phys. Rev. B {\bf 78}, 214515 (2008).

\bibitem{Cano}
A. Cano, M. Civelli, I. Eremin, and I. Paul, Phys. Rev. B {\bf 82}, 020408(R) (2010).

%\bibitem{Sun}
%D. L. Sun, J. Z. Xiao, C. T. Lin, J. of Crystal Growth {\bf 321}, 55 (2011). (Co-doped crystals)
%
%\bibitem{Lorenzana}
%J. Lorenzana, G. Seibold, C. Ortix, and M. Grilli, Phys. Rev. Lett. {\bf 101}, 186402 (2008). (competing orders in FeAs layers)%
%
%\bibitem{Barzykin}
%V. Barzykin and L. P. Gor'kov, Phys. Rev. B {\bf 79}, 134510 (2009). (role of striction at magnetic and structural transitions)
%
%\bibitem{Fernandes2}
%R. M. Fernandes, D. K. Pratt, W. Tian, J. Zarestky, A. Kreyssig, S. Nandi, M. G. Kim, A. Thaler, N. Ni, P. C. Canfield, R. J. %McQueeney, J. Schmalian, and A. I. Goldman, Phys. Rev. B {\bf 81}, 140501(R) 2010.  (unconv SC s=/- coexistence AFM-SC)
%
%\bibitem{Peercy}
%P. S. Peercy, Phys. Rev. Lett. {\bf 35}, 1581 (1975). (1st to 2nd transition tuned by atomic substution in BixSb1-xSI)
%
%\bibitem{Teng1}
%M. K. Teng, M. Massot, M. Balkanski, and S. Ziolkiewicz, Phys. Rev. B {\bf 17}, 3695 (1978) (tricritical point in BixSb1-xSI)
%
%\bibitem{Teng2}
%M. K. Teng, M. Massot, M. R. Chaves, M. H. Amaral, S. Ziolkiewicz, and W. Young, Phys. Stat. Sol. (a) {\bf 63}, 605 (1981). %(tricritical point induced by atomic substution in BixSb1-xSI -- pyroelectric coefficient dependence with T)
%
%\bibitem{Cox}
%U. J. Cox, A. Gibaud, R. A. Cowley, Phys. Rev. Lett. {\bf 61}, 982 (1988). (KMnF3... change the order of the transition by %substituting Ca for Mn -- X-ray OP)
%
\end{thebibliography}
\end{document}